# The Brahmaputra: A Socio-Political Conundrum

G. Lakshmipriya, Dr. Love Trivedi

Department of Physics, JPIS, Jaipur, Rajasthan, India

**Abstract**

The Brahmaputra mainly flows through three countries: China, India, and Bangladesh. It has been a source of conflict for the involved nations due to the various intricacies of transboundary water management such as the power play between upper and lower riparian countries, the lack of comprehensive policies for bilateral or multilateral cooperation, and the correlation of water conflicts with territorial disputes. Each country's means of safeguarding its resources through the construction of dams and hydroelectric power projects affect the other States. These conflicts have led to adverse consequences like flooding, withholding of significant meteorological data, and potential diversion of river water. This paper examines the implications of these measures on the national interests of the countries involved with respect to their national security, water security, economic development, and ecological stability. The paper has chosen to analyze these implications drawing precedence from the case studies of the Danube River and the Aswan Dam on the Nile River to deduce commonalities with the transboundary water conflict at hand. Secondly, data regarding each country's dependency on the Brahmaputra's water is interpreted to gauge the consequences of disruptions in its flow.

**Background**

Recent reports in line with China's 14th Five-Year-Plan point towards a plan to construct a hydroelectric power project that is expected to surpass the controversial Three Gorges Dam with regard to storage capacity and hydropower generation. These actions have raised concerns in India and Bangladesh due to the suspicions of water diversion by China. (Rahman, 2017)

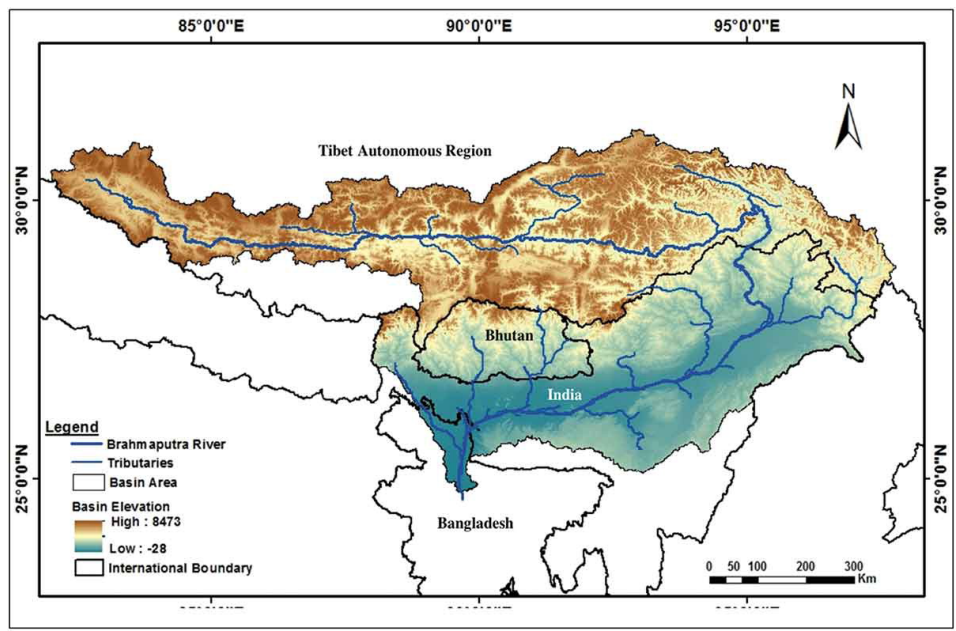

Source: (Vij & Warner, 2018)

The Brahmaputra impacts the Hindu Kush-Himalayan (HKH) region which encompasses areas of China, India, Bangladesh, Myanmar, Bhutan etc. These countries are highly dependent on the freshwater that is obtained from the river due to the water dependent nature of their economies. 90% of all water withdrawals in these countries are agriculture related. (Jiang, 2017). Out of the total catchment area 2,93,00 sq. km lies in Tibet, 2,40,000 sq km in India and Bhutan and 47,000 in Bangladesh. (Government of Assam)

**Overall Aim**

China's control over Tibet and its position as an upper riparian state confer it leverage over India and Bangladesh, which aligns with its efforts of "weaponizing water". (Ho, 2020) Housing around 20% of the world population with only 7% of water resources, China battles with an uneven and scarce distribution of water. Attempts towards alleviating the country's water scarcity crisis and meeting its hydropower targets through extensive damming of international rivers allows it to exert dominance over water resources and consequently the power and agricultural output of the affected countries.

**Strategic Impact**

China maintains an obscure behavior regarding transboundary water management and lacks a comprehensive policy for the same. The country usually prefers a bilateral approach but is often willing to participate in multilateral negotiations with smaller countries. Time and again China has been known to use the upstream strategy of rejecting multilateral frameworks such as the UN Convention on the Law of the Non-Navigational Uses of International Watercourses. It went so far as to withdraw its commissioner from the World Commission on Dams. Moreover, the country's legislation also leaves water management out of its ambit, there is only one article of the 2002 Water Law that deals with international waters. This upstream advantage coupled with the ambiguity regarding water management allows China to use the river to its advantage. For instance, in 2009, it blocked an Indian loan request from the Asian Development Bank because it was marked as a watershed project in Arunachal Pradesh. (Zhang & Li, 2018) China has been secretive about its construction of dams and seldom reveals information about future projects or long-term plans. In 2010, after years of refusal it admitted to construction of the Zangmu dam, a 510M megawatt hydropower project which raised concerns in India regarding further action on the "Great Bend" or the "Great Canyon" where the Brahmaputra curves and flows onto the Assamese plains.

**Security Impact**

**<u>Withholding of hydrological data</u>**

The sharing of hydrological data is essential for lower riparian countries for predicting floods and conducting feasibility studies. During the Doklam flash floods in 2017, China refused to provide India with date about the discharge of water in Brahmaputra. This incident was directly linked to the Doklam border standoff that damaged China-India relations. (Rahman, 2017)

**<u>Affecting water quality</u>**

In 2017, the Siang River, a tributary of Brahmaputra, blackened. Upon sampling, a test by Arunachal Pradesh's Public Health Engineering department found that the turbidity of Siang's water was substantially higher than the permissible limit. This was followed by reports of fish and aquatic animals reducing in number due to the polluted water. (Saikia, 2017)The water became unfit for use and severely affected agricultural output and fishing communities along the Siang Valley. This confirmed the extent to which China can use water resources to affect the economy of its downstream neighbors. (He, Wu, Feng, & Li, 2014)

**Linking territorial issue with water issue**

Two of China's major water projects pose a significant threat to India and Bangladesh. Firstly, the South-North Diversion Project aims to divert 44.8 billion cubic meters of water per annum to the drier regions in the Northern part of the country. This diversion is predicted to be a little more than 1,155km long and will require the construction of 23 pumping stations that will hold an installed capacity of 453.7 MW in the first of the seven stages. Moreover, China's influence over international waters can be used as a political weapon. An example of this was the collapse of an artificial dam in Tibet that claimed 26 lives and damage worth 140 crore in Arunachal Pradesh along the Siang valley.

**Economic**

Tibet holds around 200 million kWh of water resources which makes up 30% of that in China. Hydropower exploitation of 60 million kWh at the lower reaches of the Brahmaputra can provide 300 billion kWh of "clean, renewable and zero carbon" electricity annually. (Reuters, 2020) This plan of action is an effort towards meeting China's 2060 carbon neutrality goals. According to recent studies, 40% of the annual streamflow of the river comes from the catchment area in China. With 11 dams built on the Brahmaputra already, plans for water diversion will drastically affect the economic wellbeing, ecological stability and national security of India and Bangladesh.

**Global Case Studies: Dams on River Water**

The construction of mega dam projects on both the Indian and Chinese side is likely to impact the three riparian states. The Aswan Dam and Danube River case studies have been explained in order to elucidate the potential impact of future infrastructure projects on Brahmaputra.

**<u>Aswan Dam Case Study: The dam that succeeded</u>**

Built in 1970, the Aswan Dam has been one of the most controversial dams in world history for raising political, economic, and environmental issues. Much of the controversy stemmed from its construction, facing the repercussions of the Cold War, the United States withdrew its funding for the construction of the dam and as a result, it was built with assistance from the Soviet Union. It was heavily criticized by US journalists for its negative environmental impacts, especially for having been built without conducting environmental impact studies. In the 1980s and early 1990s, it was found that a substantial amount of the criticism was based on suppositions rather than facts that attempted to label the dam as a failed large-scale development project. Studies done by the Canadian International Development Agency confirmed that contrary to popular belief the Aswan Dam was not a "complete disaster" but rather "one of the best dams in the world" for the economic benefits that it provided Egypt. Comparing the ecological concerns of the Aswan Dam with that of India's proposed hydropower projects, positive light is shed on the impact of the dam. While it was predicted that the dam would cause potential drops (3-8.5 meters) in the river-bed level downstream of the dam due to erosion caused by the flow of silt-free water, it was found that the actual drop was less than 15% of the lower estimate. Moreover, the anticipated reduction in soil fertility was compensated by the addition of lime-nitrate fertilizer. This case study indicates that the effect of a dam, in some cases, may not be drastic for lower riparian countries. (Biswas, 2020)

**Danube River Case Study: An ecosystem ruined**

The Danube River is a unique case study with respect to transboundary water management. Like the Brahmaputra, the Danube River flows through different countries. Although in this case due to the length of the river it flows through 17 different countries. Similar to the Brahmaputra which significantly impacts China, India, and Bangladesh, the Danube River is of great importance to Germany, Slovakia, Austria, etc. It is considered "the single most important non-oceanic body of water in Europe". (Beckmann, 2006) The complexity of its management can be attributed to the many constructions for flood protection, agriculture, and power production. Its rich biodiversity draws global and regional attention to this region. This river is a testament to the effect of human intervention on the biodiversity of a river ecosystem. 80% of Danube's wetlands, floodplains, and forests were destroyed by man-made projects like the Danube Odra Elbe Canal, the Ukraine Danube Delta Canal, and the EU's Common Agriculture Policy, creating a straight and deep channel. Moreover, there was vast destruction to the natural environment, reducing nursery areas for spawning fish, and blocking migratory pathways. Although, unlike the Brahmaputra, cooperation on transboundary water management in the Danube River basin is governed by multilateral agreements like the Convention on Cooperation for the Protection and Sustainable Use of the Danube River. Having a multilateral approach to restoring the ecological stability of the river has proved essential for recovering from the damage caused to the biodiversity.

**The Indian Perspective**

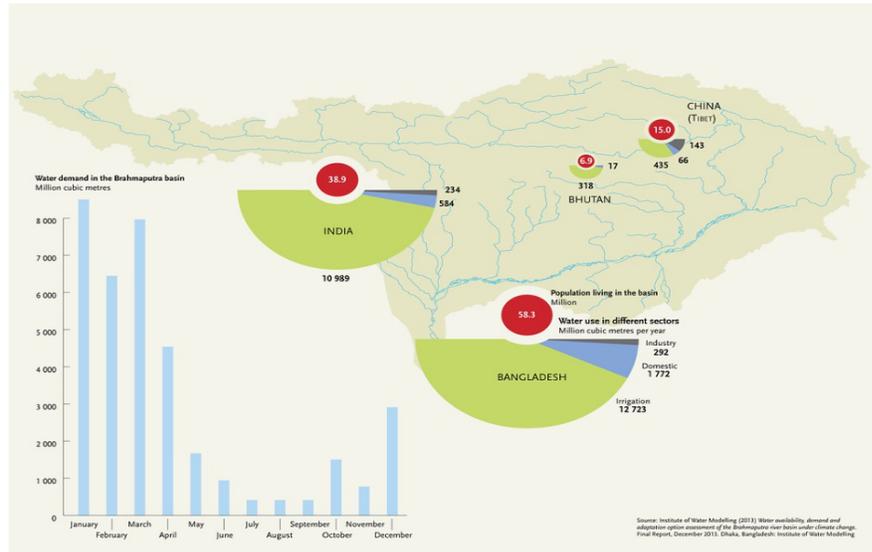

Fig 1: Water Use: Case of Brahmaputra Basin (GRID)

With 17% of the world's population residing in India, the country only has 4% of its water resources available for use. The Brahmaputra flows for 916 km and is significant for the nation's agriculture. During the river's course in the Assam Valley, it is joined by about 20 important tributaries from its North bank and 13 from its South bank that bring high sediment load. The drainage area lying in India is 194413 sq. km and this figure constitutes 5.9% of the total geographical area of the country. The regions surrounding the river have red loamy soil and alluvial soil. The culturable area of the sub basin of 12.1 M ha accounts for 6.2% of the total culturable area of the country. The tributaries of the river are rain fed and experience frequent flooding. 85% out of the total monsoon rains occur from May to September followed by dry spells. (Government of Assam)

The river is the source of 44% of all hydroelectric power generated in the country and 29% of India's freshwater resources. Further, construction of dams on the river restricts the flow of silt and lower the agricultural productivity of the region. India's main concern regarding

China's developments on the river not only include water shortages but also potential flash floods caused as a result of ineffective water management and extensive deforestation along Chinese stretches of the river. (Biba, 2013) Post the 2017 Doklam standoff, a conflict near the Sikkim border regarding China's military plans to build a road into Bhutanese territory close to India's Chicken Neck Corridor, China refused to share any meteorological data related to the river's water levels, leaving India vulnerable to floods. However, in 2018, an MOU between China and India on sharing hydrological data was signed. It required China to provide data regarding the river's levels during the flood season and also if the water level exceeds the agreed upon level during non-flood season. (Vasudeva, 2020)

**Bangladesh Perspective**

Bangladesh is the most vulnerable to this potential diversion of water since the Brahmaputra constitutes 65% of the country's river water and it is heavily dependent on the hydrological data provided by the upper riparian countries for flood prevention. The river provides 67% of the total annual river discharge of Bangladesh in the dry season, the river water supports agricultural and fishing activities. (Thakkar, 2003) Since approximately 50% of the country's population is primarily involved in agriculture and more than 70% of its land is used for crop production, the Brahmaputra River plays an integral role in sustaining in its economic activities. (Food and Agriculture Organization of the UN). In order to reduce the impact of flooding due to the river, the Indian government follows a "technocratic" approach involving infrastructure development such as the construction of storage structures and hydropower stations. These adaptation measures and the resultant damming of the river would impact the lives of people living in the lower riparian country. (Vij & Warner, 2018)

Inadequate data sharing between India and Bangladesh restricts basin level research into flooding and erosion issues which is a major threat to both countries. The sharing of meteorological data between them is only outlined in one memorandum of understanding via the annual Joint Rivers Commission meeting. (Barua & Vij, 2018)Bangladesh's position regarding transboundary water management is further complicated by its close economic ties with China and its involvement in the Belt and Road Initiative. (Ahmad, 2020) During the dry season, the southernmost riparian faces water shortages since it does not receive enough water from the Teesta, especially for rice farming. Researchers have attributed the country's rising food insecurity to its inadequate water supply.

**Conclusion**

Questions remain over the hydropower dam's social, ecological, and environmental impacts, and Beijing's planned dam on the Brahmaputra could be another sore point between the neighbors. There are concerns from downstream countries, especially India, over the dam's potential ecological and environmental damage to the Brahmaputra. Some Indian politicians believe that China's damming of the Brahmaputra could result in challenges to food security and water security in India. Over the past decade, water insecurity and water scarcity have exacerbated the frequency of South Asian droughts, which occur for longer periods, affecting more people and industries. Apart from concerns over water supply, some are wary of the geopolitical implications of China's new super-dam. As China extends its sphere of influence in Asia, there are fears in the downstream region that China could use water as a powerful political weapon to pressure the downstream region into submission over other issues, such as disputes over the Belt and Road Initiative or COVID-19. They fear that China could weaponize water by using dams to control the water flow of major transnational rivers, intentionally causing floods or water scarcity if China refuses to release any water to the downstream nations.

India plans to build a dam on the Brahmaputra in Yingkiong, Arunachal Pradesh in order to mitigate floods and ensure water security as a countermeasure against China's actions. This new dam is projected to store around 10 billion cubic. m of water. (Chaudhary, 2020)